\documentclass[aps,pra,nobalancelastpage,superscriptaddress,twocolumn,10pt]{revtex4-2}

\usepackage{color,ifthen,amsthm,amsmath,amsxtra,amsfonts,dsfont,graphicx,bm,tikz,scalerel,wasysym,bbm,graphicx,amsthm, braket,physics,empheq, comment}
\usepackage[colorlinks=true,linkcolor=blue, citecolor=blue, urlcolor=blue, bookmarks]{hyperref}
\usepackage[inline]{enumitem}

\newcommand{\be}{\begin{equation}}
\newcommand{\ee}{\end{equation}}
\newcommand{\bea}{\begin{eqnarray}}
\newcommand{\eea}{\end{eqnarray}}
\newcommand{\bse}{\begin{subequations}}
\newcommand{\ese}{\end{subequations}}
\def\rhoAQ{\rho_{A,Q}}

\theoremstyle{plain}

\theoremstyle{plain}

\theoremstyle{plain}

\usepackage{color,ifthen,amsthm,amsmath,amsxtra,amsfonts,dsfont,graphicx,bm,tikz,scalerel,wasysym, physics, bbm,  graphicx}
\usepackage[colorlinks=true,linkcolor=blue, citecolor=blue, urlcolor=blue, bookmarks]{hyperref}

\usepackage{amsthm}

\begin{document}

\title{Microscopic origin of the quantum Mpemba effect in integrable systems}

\author{Colin Rylands}
\affiliation{SISSA and INFN Sezione di Trieste, via Bonomea 265, 34136 Trieste, Italy}

\author{Katja Klobas}
\affiliation{School of Physics and Astronomy, University of Nottingham, Nottingham, NG7 2RD, UK}
\affiliation{Centre for the Mathematics and Theoretical Physics of Quantum Non-Equilibrium Systems, University of Nottingham, Nottingham, NG7 2RD, UK}

\author{Filiberto Ares}
\affiliation{SISSA and INFN Sezione di Trieste, via Bonomea 265, 34136 Trieste, Italy}

\author{Pasquale Calabrese}
\affiliation{SISSA and INFN Sezione di Trieste, via Bonomea 265, 34136 Trieste, Italy}
\affiliation{International Centre for Theoretical Physics (ICTP), Strada Costiera 11, 34151 Trieste, Italy}

\author{Sara Murciano}

\affiliation{Walter Burke Institute for Theoretical Physics,  Caltech,  Pasadena CA 91125, USA}
\affiliation{Department of Physics and IQIM, Caltech, Pasadena, CA 91125, USA}

\author{Bruno Bertini}
\affiliation{School of Physics and Astronomy, University of Nottingham, Nottingham, NG7 2RD, UK}
\affiliation{Centre for the Mathematics and Theoretical Physics of Quantum Non-Equilibrium Systems, University of Nottingham, Nottingham, NG7 2RD, UK}

\begin{abstract}
  The highly complicated nature of far from equilibrium systems  can lead to a complete breakdown of the physical intuition developed in equilibrium.  A famous example of this is the Mpemba effect which states that non-equilibrium states may relax faster when they are further from equilibrium or,  put another way,  hot water can freeze faster than warm water.  Despite possessing a storied history, the precise criteria and mechanisms underpinning this phenomenon are still not known. Here we study a quantum version of the Mpemba effect that takes place in closed many body systems with a $U(1)$ conserved charge: in certain cases a more asymmetric initial configuration relaxes and restores the symmetry faster than a more symmetric one. In contrast to the classical case, we establish the criteria for this to occur in arbitrary integrable quantum systems using the recently introduced entanglement asymmetry. We describe the quantum Mpemba effect in such systems and relate properties of the initial state, specifically its charge fluctuations, to the criteria for its occurrence.  These criteria are expounded using exact analytic and numerical techniques in several examples, a free fermion model, the Rule 54 cellular automaton, and the Lieb-Liniger model. 
\end{abstract}

\maketitle
\textit{Introduction.---}
Consider a system in thermal equilibrium with a bath.  If we wish to lower the temperature of the system we can gradually change the temperature of the bath to the desired final value with the system following suit.  Naturally, the time taken for this process depends on the temperature difference: the greater the initial temperature the longer the time to effect the change.  Therefore, it should come as a surprise that if instead the bath temperature is quenched, i.e. changed suddenly, the opposite can happen,  an initially hotter system can cool faster.  This counter-intuitive phenomenon, equally likely to be found in the lab or at home, is known as the Mpemba effect~\cite{mpemba1969cool}. Originally described in the context of freezing water, the effect  boasts a long history stretching back millennia and has since been extended to encompass a number of other systems~\cite{ahn2016experimental,hu2018confirmation, lasantaMpemba, klich2019mpemba,kumar2020exponentially,kumar2022anomalous, walker2023Mpemba, tyr-23, walker2023optimal, bera2023effect}.  Despite this, however, debate remains over its precise origin and predictability~\cite{burridge2016questioning,lu2017nonequilibrium}.  A general criterion for its occurrence and a full understanding of the phenomenon remain elusive.   

In this paper, we investigate such anomalous relaxation dynamics of a non-equilibrium state,  not in the context of a classical, open system as described above but, instead, using an isolated, many-body quantum system possessing a conserved $U(1)$ charge. In this context a natural analogue of the Mpemba effect is observed when an initial configuration that is more charge asymmetric, and hence more out-of-equilibrium, relaxes faster than a more symmetric one~\cite{ares2022entanglement}. 
 We note that Mpemba effects have been studied previously in quantum systems, however, these have used thermal states and/or open systems~\cite{nf-19, cll-21, kck-22, manikandan-21, ias-23, chatterjee2023quantum, shapira2024mpemba,zhang2024observation}.  In contrast, herein we study the quench dynamics of zero temperature states in a closed system with the effect being driven instead by quantum rather than thermal fluctuations. We establish two criteria for its occurrence in integrable systems, which are then expressed in terms of the charge probability distribution of the initial state. This, therefore, imbues the quantum Mpemba effect with a level of predictability absent in the classical counterpart. We illustrate our treatment through a number of representative examples including both free and interacting models as well as cases where the Mpemba effect is present and absent. 

\begin{figure*}[t]
\centering
\includegraphics[width=0.48\textwidth]{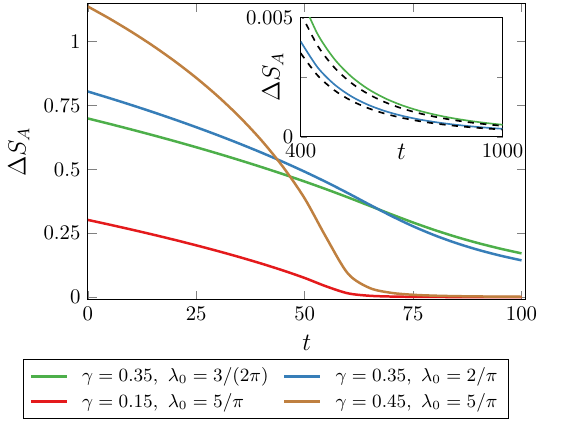}
\includegraphics[width=0.48\textwidth]{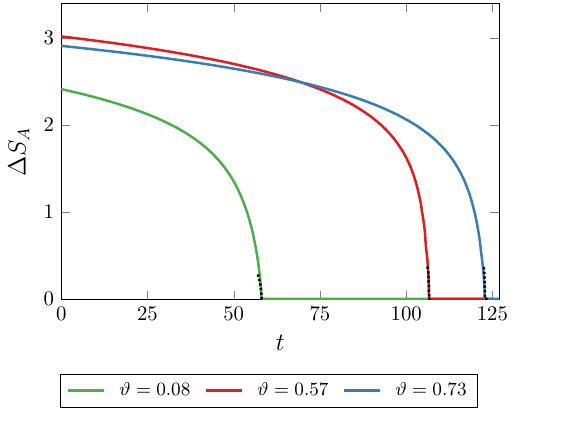}
  \caption{\label{fig:qme}
    Evolution of the entanglement asymmetry $\Delta S_A$ as a function of time for a subsystem of $\ell=100$ sites. Different curves correspond to different initial states and the {\it crossings signal the occurrence of QME}. The figure reports the results for a free fermionic system (left panel) and for the interacting Rule 54 quantum cellular automaton (right panel). The results are obtained using Eq.~\eqref{eq:AsymmetryGeneral1}. {The value of the parameters that determine the occurrence of QME, see main text, are for the curves in the left panel $(\mathcal{X}, \vartheta'')=(6\times 10^{-3}, 0.55),( 7\times 10^{-3}, 0.35), ( 1.4\times 10^{-2}, 8.1\times 10^{-4}), (1.7\times 10^{-3}, 9\times 10^{-5})$ (in clockwise order according to the legend) and, in the right panel, $\mathcal{X}=0.07, 0.25, 0.20$ (from left to right in the legend). }
    }
\end{figure*}

\textit{Setting.---}
We consider a closed quantum system whose dynamics are governed by an integrable Hamiltonian $H$ and which possesses a $U(1)$ conserved  charge $Q$, i.e.  $[H,Q]=0$.  The system is initialized in a  broken symmetry state $\rho=\ketbra{\Psi_0}$ such that $[\rho,Q]\neq 0$, which is not an eigenstate of $H$, $[\rho,H]\neq 0$, and then allowed to evolve unitarily according to $e^{-i Ht}$. Since the total system is closed, $\rho(t)$ will never relax to a stationary state. Instead, we study a subsystem $A$ of length $\ell$ described by the reduced density matrix $\rho_A(t)=\text{tr}_{\bar{A}}[\rho(t)]$ with $\bar{A}$ being the complement of $A$. We denote the restriction of $Q$ to the subsystem by $Q_A$, and its average by $q_0=\tr[\rho_A(0)Q_A]$. This restriction to the quench dynamics of a subsystem results in nontrivial behavior of the charge fluctuations throughout $A$ due to the broken $U(1)$ symmetry in the initial state. Under generic circumstances, however, we expect that locally the system relaxes to a stationary state~\cite{PolkovnikovReview, calabrese2016introduction, VidmarRigol, essler2016quench, doyon2020lecture, bastianello2022introduction, alba2021generalized}, and, barring some exotic instances~\cite{fagotti2014on, bertini2015pre, ares2023lack}, that this stationary state restores the $U(1)$ symmetry, i.e., $\lim_{t\to\infty}[\rho_A(t),Q_A]=0$. This symmetry restoration can therefore serve as a proxy, albeit a coarse-grained one, of the relaxation to the stationary state and thus allows us to study the quantum Mpemba effect (QME) in a closed quantum system. One can compare different broken symmetry states and determine if states with ``more'' symmetry breaking can relax faster. 

To quantify the notion of symmetry breaking and restoration 
we use the entanglement asymmetry, $\Delta S_A(t)$, a recently introduced measure which distills the interplay between the spreading of entanglement and the dynamics of 
a conserved charge~\cite{ares2022entanglement,ares2023lack,ferro2023nonequilibrium, capizzi2023entanglement, capizzi2023universal,caceffo2024entangled,joshi2024observing}. The entanglement asymmetry is defined as the relative entropy between two different reduced density matrices: $\rho_A(t)$ and its symmetrized version $\rhoAQ=\sum_q \Pi_q \rho_A(t)\Pi_q$ where $\Pi_q$ is the projector on to the eigenspace of $Q_A$ with eigenvalue $q$, namely 
\begin{equation}\label{eq:AsymmetryDef}
\Delta S_A(t)=\tr \left[\rho_{A}(t)\left(\log\rho_A(t)-\log\rhoAQ(t)\right)\right].
\end{equation}
Being a relative entropy, $\Delta S_A(t)\geq 0$, and it can be shown that equality is achieved only when $[\rho_A(t),Q_A]=0$, that is when the symmetry is restored locally. Given the expected relaxation of the state at long times this translates to $\lim_{t\to\infty}\Delta S_{A}(t)=0$.  Using $\Delta S_{A}(t)$ we can quantify how far a state is from being symmetric and, therefore, also how close it is to the stationary state. For instance, for two states $\rho_{A,j}$, $j=1,2$ we say that the symmetry is broken more in $\rho_{A,1}$ than $\rho_{A,2}$ if $\Delta S_{A,1}>\Delta S_{A,2}$ and accordingly the former is further from equilibrium than the latter. 

With this in mind the QME is defined as occurring if two conditions are met
\begin{enumerate}[label=(\roman{enumi})]
  \item\label{it:cond1}
    $\quad\Delta S_{A,1}(0)-\Delta S_{A,2}(0)>0$,
  \item\label{it:cond2}
    $\quad\Delta S_{A,1}(\tau)-\Delta S_{A,2}(\tau)<0, \qquad \forall\tau>t_\text{M}$,
\end{enumerate}
where $t_{\text{M}}$ is the Mpemba time. That is,  while $\rho_{A,1}$ is initially further from equilibrium than $\rho_{A,2}$ it relaxes to the stationary state faster. Fig.~\ref{fig:qme} shows examples of this phenomenon occurring (or not) in both a free and an interacting system.

\textit{General treatment.---}
We aim to compute the entanglement asymmetry and relate conditions~\ref{it:cond1} and~\ref{it:cond2} to properties of the initial state $\ket{\Psi_0}$ thereby obtaining a predictive criteria for the QME. To this end, we start by employing the replica trick and defining the R\'enyi entanglement asymmetry as
\begin{equation}\label{eq:Renyi_asymm}
  \Delta S_{A}^{(n)}=\frac{1}{1-n}\left(\log\tr(\rhoAQ^n)-\log\tr(\rho^n_{A})\right),
\end{equation}
whose limit $n\to 1$ gives~\eqref{eq:AsymmetryDef}. Using the Fourier representation of the projector {$\Pi_q=\int_{-\pi}^\pi\frac{\rm{d}\alpha}{2\pi}e^{i\alpha(Q_A-q)}$}, the moments of $\rhoAQ$ are
\begin{equation}\label{eq:FT}
 \tr\left[\rhoAQ^n(t)\right]=\int_{-\pi}^\pi \frac{{\rm d}\boldsymbol{\alpha}}{(2\pi)^{n-1}} \delta_p\bigl({\textstyle \sum_{j}}\alpha_j\bigr) Z_n(\boldsymbol{\alpha}, t),
 \end{equation}
where $ Z_n(\boldsymbol{\alpha}, t)=
\tr\left[\prod_{j=1}^n(\rho_A(t) e^{i\alpha_{j} Q_A})\right]$ are the charged moments and $\delta_p(x)$ is the $2\pi$-periodic Dirac delta function. 
This form is particularly useful, as the charged moments can be efficiently calculated: both for interacting~\cite{bertini2022nonequilibrium,bertini2023dynamics} and non-interacting dynamics~\cite{ares2022entanglement}. In particular, as we show in the SM \cite{Note1}, these quantities take a particularly simple form for ${n\to1}$, which is our limit of interest. For clarity we focus on the case of a single quasiparticle species, although the extension to many is straightforward. In this case, using an emergent quasiparticle picture we find~\cite{Note1, Note11}
\begin{equation}\label{eq:charged_1}
  \frac{\tr[\rhoAQ^n]}{\tr[\rho_{A}^n]}=I_n+\mathcal{O}\left[(n-1)^2\right],
\end{equation}
with the definition
\begin{equation}
    \!\!\!\!\!\!\!\! I_n\!=\!\!\int\limits_{-\pi}^{\pi}
    \!\!\!\frac{{\rm d}\boldsymbol{\alpha}}{(2\pi)^{n-1}} 
    \delta\bigl({\textstyle \sum_{j}}\alpha_j\bigr) e^{ \ell \sum_{j=1}^{n} \int \!\!\mathrm{d}\lambda\, x_\zeta(\lambda)f_{\alpha_{j}}(\lambda)}\!,
    \label{eq:In}
\end{equation}
where $x_\zeta(\lambda)=\max[1-2 |v(\lambda)| \zeta ,0]$,
$v(\lambda)$ is the quasi-particle velocity, $\zeta=t/\ell$ is the rescaled time variable, and $f_{\alpha}(\lambda)$ specifies the contribution of the mode $\lambda$. The precise form of $f_\alpha$ is unimportant but we note that it satisfies $f^*_\alpha(\lambda)=f_{-\alpha}(\lambda)$ and $f_{\alpha}(\lambda) = f_{\alpha+\pi}(\lambda)$. By using the Poisson summation formula~\cite{pinsky2008introduction} to rewrite the periodic delta function, we obtain the  series representation of $I_n$
\begin{equation}
\label{eq:defJk}
   I_n \!\!=\!\! \smashoperator{\sum_{k=-\infty}^{\infty}} \!J_k^n,
\quad 
  J_k\!=\!\!\int\limits_{-\pi}^{\pi} \!\!\frac{\mathrm{d}\alpha}{2\pi}
    e^{i k \alpha}  \exp[\ell\!\!\int \!\!\mathrm{d}\lambda\, x_\zeta(\lambda) f_{\alpha}(\lambda)].
\end{equation}
To express the asymmetry $\Delta S_A$, we need to perform the analytic
continuation $n\to {z\in\mathbb{C}}$. Since the resulting function has to be real for $z\in\mathbb{R}$, we make the minimal choice
\begin{equation}
  I_n \to I_z,\qquad
  J_k^n \to \frac{1}{2}\left(e^{z \log J_k}+e^{z\log J_k^{\ast}}\right).
  \label{eq:analyticcont}
\end{equation}
With this choice we obtain our first main result: a completely general expression
for the entanglement asymmetry
\begin{equation}
\label{eq:AsymmetryGeneral1}
  \Delta S_A=-\lim_{z\to 1} \left[\partial_z I_z\right]=
  - \smashoperator{\sum_{k=-\infty}^{\infty}}
  \Re[J_k(t) \log J_k(t)].
\end{equation}

This formulation allows us to rewrite the conditions~\ref{it:cond1} and~\ref{it:cond2} stated above in terms of properties of the initial state. Namely, by expanding $\Delta S_{A}(t)$ for large $\ell$ both at $t=0$ and $t\gg \ell$ we can express the QME conditions as~\cite{Note1} 
\begin{enumerate}[label=(\roman{enumi})]
  \item\label{it:ncond1}
    $\quad J_{q_{0,1},1}(0)-J_{q_{0,2},2}(0)<0$,
  \item\label{it:ncond2}
    $\quad J_{0,1}(t)-J_{0,2}(t)>0,\qquad t\gg \ell$,
\end{enumerate}
where $q_{0,j}$ denotes the expectation value of the charge in $\rho_{A,j}(0)$.
These conditions have a clean physical interpretation which we now explain.  
Originally our first condition required that $\rho_{A,1}(0)$ be more asymmetric than $\rho_{A,2}(0)$ and so it is natural to expect that the fluctuations of charge in $\rho_{A,1}(0)$ be larger than in $\rho_{A,2}(0)$. This expectation is indeed borne out: $J_{k}(0)$ can be shown~\cite{Note1} to take the following form in the leading order in $\ell$,
\begin{equation} \label{eq:Jk0gen}
  J_{k}(0)\simeq \frac{1}{\sqrt{\pi \sigma_{0}^2}} e^{-\frac{(k-q_0)^2}{2\sigma_0}},
\end{equation}
where $q_0$ is the expectation value of the charge in the initial state, and $\sigma_0$ is the variance of the charge. In other words, $J_{q_{0,j},j}(0)$ is proportional to the probability of measuring the charge equal to $q_{0,j}$ in $\rho_{A,j}(0)$. Therefore, \ref{it:ncond1} is equivalent to there being a lower probability of measuring the expectation value of the charge in the more asymmetric state. That is, the more asymmetric the state the less peaked its charge probability is about the average.  For the second condition, the original statement is that after some time the asymmetry --- and therefore the charge fluctuations --- are greater in $\rho_{A,2}(t)$ than $\rho_{A,1}(t)$. However, the only way for the subsystem to suppress its charge fluctuations is by transporting charge through its boundaries.  Thus one can expect that in the more asymmetric case the charge fluctuations are transported predominantly by the faster modes and hence allow for faster relaxation.  This is  encapsulated in the rewritten condition. One can interpret $J_{0,j}(t)$ at long times, $t\gg\ell$, as the probability that slowest modes transport no charge, and hence produce no charge fluctuations. To see this we note that $J_0(0)/\sqrt{2}$ gives the probability of measuring charge $q=0$ in the initial state (cf.\ Eq.~\eqref{eq:Jk0gen}). For $t\neq 0$ however the presence of $x_\zeta(\lambda)$ in Eq.~\eqref{eq:defJk} has the effect of filtering out the contribution of fast modes. In particular, for large times we have 
\be
\exp[\ell \int \!\!\mathrm{d}\lambda\, x_\zeta(\lambda) f_{\alpha}(\lambda)] \approx \tr[\rho_A e^{i \alpha Q_{\rm sl}}],
\ee 
where $Q_{\rm sl}$ is the charge of the slow modes. Accordingly, since the slowest modes are the last to relax, the charge distribution relaxes faster if they are less involved in the charge transport.  

Combining these conditions, we can state that the QME occurs for a class of initial states if, by increasing the charge fluctuations in a subsystem, one simultaneously has the slowest modes carrying less charge and thereby resulting in faster charge transport. The above discussion demystifies the QME, connecting it with the charge distribution of the initial state among the transport modes and constitutes the second main result of this paper.

\textit{Examples.---}
In the following, we will illustrate and test our predictions in three concrete examples, which are ordered according to the amount of explicit results we can obtain. In particular, we show how our results allow to predict all the crossings in Fig.~\ref{fig:qme}. 

\noindent\underline{Example 1}
Let us first consider the simplest possible setting, i.e., free fermionic systems. For concreteness, we focus on a simple model with Hamiltonian $H=\int_{-\pi}^\pi {\rm d}\lambda \,\epsilon(\lambda) \eta_\lambda^\dagger \eta_\lambda,$
where $\eta_\lambda^\dagger$ and $\eta_\lambda$ are canonical fermionic creation and annihilation  operators and $\epsilon(\lambda)$ is the single-particle dispersion relation.  $H$ has a single $U(1)$ conserved charge, the total particle number, 
$Q=\int {\rm d}\lambda\,\eta_\lambda^\dagger \eta_\lambda$.  The initial configuration is given by a squeezed state,  which breaks this symmetry and can be written in terms of eigenstates of the post-quench Hamiltonian as $\ket{\Psi_0}={\rm exp}\left[-\int_0^{\pi} {\rm d}\lambda \mathcal{M}(\lambda)\eta_\lambda^\dagger \eta_{-\lambda}^\dagger\right]\ket{0}.$ Here  $\mathcal{M}(\lambda)$ is an arbitrary real and odd function and $\eta_\lambda\ket{0}=0$ for all $\lambda$.
In this case, Eq.~\eqref{eq:charged_1} applies exactly for any $n\in\mathbb{N}$ with $v(\lambda)= \epsilon'(\lambda)$ and $f_{\alpha}(\lambda) =\log \left(1-\vartheta(\lambda)+\vartheta(\lambda)e^{2i\alpha}\right)/(4\pi)$, where $\vartheta(\lambda)=\mathcal{M}(\lambda)^2/(1+\mathcal{M}(\lambda)^2)$ is the quasiparticle occupation function. Given the simple form of $f_{\alpha}(\lambda)$, the integral over $\alpha$ in Eq.~\eqref{eq:defJk} can be evaluated (see Sec.~\ref{sec:FFgenericn} of the SM~\cite{Note1}), the analytic continuation to take the limit $n\to 1$ is unambiguous and the full time evolution of $\Delta S_A$ can be studied with Eq.~\eqref{eq:AsymmetryGeneral1}. The latter is computed numerically by truncating the series to produce the plots in Fig.~\ref{fig:qme}a and can be used to verify the asymptotic expansions used to obtain \ref{it:ncond1} and \ref{it:ncond2} (explicitly reported with a dotted line in the figures' inset).
In particular, for the figure we chose $\epsilon(\lambda)=\sqrt{\lambda^2+1}$ and $\mathcal{M}(\lambda)= \gamma \big(g(\lambda+\lambda_0)+g(-\lambda-\lambda_0)-g(\lambda-\lambda_0)-g(-\lambda+\lambda_0)\big)$ with $g(x)=1/(1+e^{-4(x - 1/\pi)})$.

  The conditions for QME are easily specialised to the non-interacting case.
Specifically, we have $2\sigma_0^2/\ell=\mathcal X \equiv \int_{-\pi}^{\pi}{\rm d}\lambda\chi(\lambda)$ 
where 
$
\chi(\lambda)=\vartheta(\lambda)(1-\vartheta(\lambda))/(2\pi)
$
is the charge susceptibility per mode. Given the specific form of $v(\lambda)$ and $\mathcal{M}(\lambda)$ 
we obtain
\begin{equation}\label{eq:asymptmt}
  J_0\simeq 1-\frac{\vartheta''(0)}{48\pi}\ell \Lambda_\zeta^3,
\end{equation}
where $\Lambda_\zeta=1/(2\zeta v'(0))$ determines the window of the slowest modes, i.e. $x_{\zeta}(\lambda)\neq 0 $ for $\lambda\in [-\Lambda_\zeta,\Lambda_\zeta]$. This means that QME occurs if $\mathcal X_1 >\mathcal X_2$ and $\vartheta''_2(0)>\vartheta''_1(0)$.\\

\noindent\underline{Example 2} Let us now move on to consider what can be regarded as the natural next step from free systems, i.e., the quantum cellular automaton Rule 54~\cite{bobenko1993two}. This is a locally interacting chain of $L$ spin-$1/2$ variables (or qubits) where the time evolution happens in discrete time steps. Rule 54 is Bethe ansatz integrable~\cite{gombor2022integrable,friedman2019integrable} and, therefore, supports stable quasiparticles. Although these quasiparticles undergo non-trivial elastic scattering, the latter is simple enough to allow a wealth of explicit results that are not available for ``generic'' interacting integrable models~\cite{prosen2016integrability,prosen2017exact,gopalakrishnan2018operator,gopalakrishnan2018hydrodynamics,inoue2018two,friedman2019integrable,alba2019operator,klobas2019time,buca2019exact,alba2021diffusion,klobas2020matrix,klobas2020space,klobas2021exact,klobas2021exactrelaxation,klobas2021entanglement}, see Ref.~\cite{buca2021rule} for a recent review. 

As time is discrete, the dynamics are generated by the evolution operator for a time-step rather than by a Hamiltonian. Alternatively, one can think of it as the Floquet operator for a periodically driven system in continuous time. Its explicit form reads as ${{\mathbb{U}}={\Pi}^{\dagger} {\mathbb{U}}_{\mathrm{e}} {\Pi} \hat{\mathbb{U}}_{\mathrm{e}}}$, where $\Pi$ is the periodic one-site shift and ${\mathbb{U}}_{\mathrm{e}}= \prod_{j\in \mathbb{Z}_L} {U}_{2j}$.
The local operator ${U}_j$ acts non-trivially only on site $j$, implementing a deterministic update that depends on the states of the neighbouring sites, i.e., $\mel{s_1^{\prime} s_2^{\prime} s_3^{\prime}}{{U}_2}{s_1^{\phantom{\prime}} s_2^{\phantom{\prime}} s_3^{\phantom{\prime}}}
 \!\! =\!\!
  \delta_{s_1^{{\prime}},s_1}
  \delta_{s_2^{{\prime}},s_1+s_2+s_3\!\!\pmod{2}}
  \delta_{s_3^{{\prime}},s_3}$.

Remarkably, this system admits a class of \emph{solvable} states with a completely accessible quench dynamics written as $\ket*{\Psi_0}=(\sqrt{1-\vartheta}\ket{00}+\sqrt{\vartheta}\ket{01})^{\otimes L/2}$~\cite{klobas2021exactrelaxation}. Here $\{\ket{0}, \ket{1}\}$ is the computational basis for one qubit and the parameter ${0<\vartheta<1}$ fixes the quasiparticle occupations. Specifically, recalling the quasiparticles in Rule 54 are specified by a ``discrete rapidity'' $\nu=\pm$~\cite{friedman2019integrable}, after a quench from a solvable state we have $\vartheta_+=\vartheta_-=\vartheta$~\cite{klobas2021exactrelaxation}.  

The solvable states are not eigenstates of the $U(1)$ charge $Q=\sum_j  (1-\sigma_{j+1}^z)\left(2+\sigma_{j+2}^z+\sigma_j^z\sigma_{j+2}^z\right)/4$, therefore, one can study the restoration of the latter, and the possible emergence of QME, by computing the entanglement asymmetry. In particular, employing a quasiparticle picture we find that Eq.~\eqref{eq:charged_1} holds in Rule 54 provided that the integral over $\lambda$ is replaced by the sum over the discrete rapidities $\nu$~\cite{Note1}. In this case we obtain $f_{\alpha,\nu} =\log (1-\vartheta+\vartheta e^{2i\alpha})/2$. Performing then the analytic continuation in Eq.~\eqref{eq:analyticcont} we arrive at Eq.~\eqref{eq:AsymmetryGeneral1} where, however, $J_k$ are computed explicitly as   
\begin{equation}
\label{eq:JkR54}
J_k = \begin{cases} 
\displaystyle
  \binom{\ell x_\zeta}{k/2}\bar{\vartheta}^{k/2}(1-\bar{\vartheta})^{\ell x_\zeta -k/2} & k/2 \in \mathbb N\\
  0 & {\rm otherwise}\\
  \end{cases}\,.
\end{equation}
Here $\bar \vartheta = {\rm min}(\vartheta,1-\vartheta)$ and $x_\zeta=\max[1-2 v_\vartheta \zeta ,0]$ with $v_\vartheta=1/(1+2\vartheta)$. Plugging this explicit expression in Eq.~\eqref{eq:AsymmetryGeneral1} we can easily compute the full time evolution of the asymmetry: some representative examples are reported in Fig.~\ref{fig:qme}b. This can be again used to check that our asymptotic analysis agrees with the exact values (the asymptotic predictions of SM~\cite{Note1} are explicitly shown by the dotted lines).

Moreover, combining~\eqref{eq:JkR54} with~\ref{it:ncond1}, and~\ref{it:ncond2}, we can again find simple conditions for the occurrence of QME.
 In particular, we have that the variance in Eq.~\eqref{eq:Jk0gen} is again written in terms of the charge susceptibility per mode, i.e., $2\sigma_0^2/\ell=\mathcal{X}\equiv\vartheta(1-\vartheta)$ while Eq.~\eqref{eq:JkR54} gives $J_0 \simeq 1+\ell x_\zeta \log(1-\bar \vartheta)$. A sufficient condition for QME is then $\mathcal{X}_1>\mathcal{X}_2$ and $v_{\vartheta_1} > v_{\vartheta_2}$. Recalling the definition of the effective velocity the second condition becomes ${\vartheta_1} < {\vartheta_2}$.

Note that in the case of Rule 54 one can independently test the validity of the asymptotic analysis but not of the analytic continuation Eq.~\eqref{eq:analyticcont}. Indeed, the ratios ${\tr[\rhoAQ^n]}/{\tr[\rho_{A}^n]}$ are only accessible for $\zeta \leq 1/2$ and $\zeta \geq 3/2$ for generic $n$~\cite{klobas2023inprep} and one cannot use the quasiparticle picture to interpolate between them~\cite{bertini2022growth}.\\

\noindent\underline{Example 3} Lastly, we consider the Lieb-Linger model of interacting bosons. The Hamiltonian is given by $H=\int_0^L {\rm d}x\, b^\dag(x)[-\partial_x]^2b(x)+c[b^\dag(x)b(x)]^2$ where $b^\dag(x), b(x)$ obey canonical bosonic commutation relations and we take $c>0$. 
Again it has a single $U(1)$ charge, the particle number, $Q=\int b^\dag(x)b(x)$. 
We quench the system from a coherent state which breaks this symmetry, i.e., $\ket{\Psi_0}=e^{-\frac{1}{2}Ld+\sqrt{d} b_0}\ket{0}$ where $b_0=\int_0^L {\rm d}x\, b^\dag(x)$, $\ket{0}$ is the vacuum and $d$ is the average charge density. 
The model is integrable and many analytic formulae for the quench from this state have been previously obtained~\cite{denardis2014solution} and are collected in the SM along with the expression for $f_\alpha$~\cite{Note1}. 
We find that $2\sigma_0^2=\ell d$, and that the QME always occurs between states with different values of $d$. Indeed, to check condition~\ref{it:ncond2} we perform an analysis of $J_0$ which at long times has a form similar to the free case~\cite{Note1}
\be
J_0(\mu)\propto 1-\frac{ \ell  \Lambda_\zeta^3 }{24 }|\vartheta''(0)|\rho^t(0),
\ee
where $\rho^t(0)$ is the density of states of the slowest quasiparticles.
Unlike the free case, $\Lambda_\zeta$ acquires a state dependence via the interactions and the behavior is governed by $|\vartheta''(0)|$ which is a rapidly decreasing function of $d$  thus verifying that condition~\ref{it:ncond2} holds at long times. 

\textit{Conclusions.---}
In this letter we provided a microscopic characterisation of the quantum Mpemba effect in integrable models by linking it to the properties of the charge distribution in the initial state. Namely, we showed that the effect occurs for two initial states $\ket{\Psi_1}$ and $\ket{\Psi_2}$ if the first state has a broader charge distribution, i.e. is more asymmetric, but stores less charge in the slower modes, making the charge transport faster. 

In fact, the connection that we established between QME and the transport properties of the system should not rely on integrability, since the QME does not require the latter to occur~\cite{ares2022entanglement,joshi2024observing}. In the presence of weak integrability breaking terms resulting in a finite quasiparticle lifetime, or in chaotic systems which have no quasiparticles, one could still relate the QME to a faster transport of charge. More generally, relating the anomalous relaxation to transport properties could also prove to be a fruitful direction in the characterisation of the classical Mpemba effect.

\begin{acknowledgments}
This work has been supported by the Royal Society through the University Research Fellowship No.\ 201101 (B.\ B.), the Leverhulme Trust through the Early Career Fellowship No.\ ECF-2022-324 (K.\ K.), the European Research Council under Consolidator grant number 771536 ``NEMO'' (F.\ A., P.\ C.\ and C.\ R.), the Caltech Institute for Quantum Information and Matter (S.\ M.), and the Walter Burke Institute for Theoretical Physics at Caltech (S.\ M.).
B.\ B., P.\ C., K.\ K., and S.\ M.\ warmly acknowledge the hospitality of the Simons Center for Geometry and Physics during the program ``Fluctuations, Entanglements, and Chaos: Exact Results'' where this work was completed.
\end{acknowledgments}

\bibliographystyle{apsrev4-2}
\bibliography{MpembaBib.bib}  

\footnotetext[1]{See the Supplemental Material which includes reference \cite{carlson}.}
\footnotetext[11]{Note that for non-interacting systems the expression above is exact to all orders in $n-1$}

\onecolumngrid
\newcounter{equationSM}
\newcounter{figureSM}
\newcounter{tableSM}
\stepcounter{equationSM}
\setcounter{equation}{0}
\setcounter{figure}{0}
\setcounter{table}{0}
\setcounter{section}{0}
\makeatletter
\renewcommand{\theequation}{\textsc{sm}-\arabic{equation}}
\renewcommand{\thefigure}{\textsc{sm}-\arabic{figure}}
\renewcommand{\thetable}{\textsc{sm}-\arabic{table}}

\begin{center}
  {\large{\bf Supplemental Material for\\ ``Microscopic origin of the quantum
  Mpemba effect in integrable systems''}}
\end{center}
Here we report some useful information complementing the main text. In particular
\begin{itemize}
  \item[-] In Sec.~\ref{sec:AsyAsymmetryGeneral1} we perform the asymptotic analysis of Eq.~\eqref{eq:AsymmetryGeneral1}.
  \item[-] In Sec.~\ref{sec:ExpansionInN} we report a proof of Eq.~\eqref{eq:charged_1} of the main text in three different cases. Free fermions, Rule 54, and generic interacting integrable models. 
   \item[-] In Sec.~\ref{sec:FFgenericn} we consider an asymptotic expansion of $\Delta S_A^{(n)}$ for $n\geq 2$ in free fermionic systems. The analytical continuation of the final result is unique and justifies our choice \eqref{eq:analyticcont} for the analytic continuation.
    \item[-] In Sec.~\ref{sec:LLdetails} we provide details and explicit formulae for the quench of the Lieb Liniger model discussed in the main text. 
\end{itemize}

\section{Asymptotic analysis of Eq.~(\ref{eq:AsymmetryGeneral1})}
\label{sec:AsyAsymmetryGeneral1}
Performing an asymptotic expansion of Eq.~\eqref{eq:AsymmetryGeneral1}, we can characterize the behaviour of $\Delta S_{A}(t)$ for both $t\ll \ell$, and $t\gg \ell$, and hence extract the precise conditions for the occurrence of the QME. We begin by focusing on the former.  We consider the scaling limit of large $\ell$, while keeping the ratio $\zeta=t/\ell$ fixed. In this case, the integrand in~\eqref{eq:defJk} scales with $\ell$, and $J_k$ can be evaluated by a saddle-point approximation. To this end, we note that the periodicity of $f_{\alpha}(\lambda)$ implies $J_{2k-1}=0$, while for even $k$ we obtain the Gaussian form 
\begin{equation} \label{eq:asyJk}
  J_{k}\simeq  \frac{1}{\sqrt{\pi \sigma_\zeta^2}}
e^{- \textstyle\frac{(k-k_\zeta)^2}{2 \sigma^2_\zeta}},
\end{equation}
where $\simeq$ denotes equality at leading order in $\ell$. We introduced the mean value ${k_\zeta =- i \int \mathrm{d}\lambda x_\zeta(\lambda) \dot{f}_0(\lambda)}$, the variance
\be
2\sigma_\zeta^2 = - \ell \int \mathrm{d}\lambda x_\zeta(\lambda)  \ddot{f}_0(\lambda),
\ee  
and denoted by $\dot{(\cdot)}$ derivatives with respect to $\alpha$.
{ Note that $\sigma_{0}$ is the variance of the charge distribution of the initial state, and $k_0$ coincides with the expectation value of charge in the initial state, $k_0=q_0$.} Indeed, for large $\ell$ the probability $P(q)$ of measuring a charge $q\in \mathbb Z$ on the initial state is~\cite{bertini2023dynamics}
\be\label{eq:chargeprob}
P(q)=\int_{-\pi}^\pi\frac{{\rm d}\alpha }{2 \pi}e^{-i \alpha q}Z_{1}(\alpha,0) \simeq \frac{e^{-\textstyle\frac{(q-q_0)^2}{2\sigma^2_{0}}}}{\sqrt{2\pi \sigma_{0}^2}}.
\ee
Plugging Eq.~\eqref{eq:asyJk} into Eq.~\eqref{eq:AsymmetryGeneral1}, we obtain the expression for the asymmetry in the short-time regime
\begin{equation}\label{eq:asymmetry_large_ell}
  \Delta S_A \simeq \frac{1}{2}\log\left[ \pi \sigma_\zeta^2 \right]+\frac{1}{2}.
\end{equation}
Because of the monotonicity of the $\log$, we have that condition~\ref{it:ncond1} is fulfilled if $\sigma_{0,1}$ is larger than $\sigma_{0,2}$.  Namely, very intuitively, the asymmetry is larger for state (1) if its charge distribution is less peaked around the mean value compared to that of state (2).

Let us now move to consider the second limit of interest, i.e., $t\gg \ell$. To treat this case we note that if $t$ is sufficiently large, $x_\zeta(\lambda)$ becomes zero everywhere, except for a finite interval around the point $\lambda^{\ast}$ where the magnitude of the velocity is minimal (in most cases, $v(\lambda^{\ast})=0$). Therefore, the scaling of the integral in the exponent of Eq.~\eqref{eq:defJk} depends on the behaviour of $f_{\alpha}(\lambda)$ close to $\lambda^{\ast}$. Specifically, the integral can be conveniently split into
\begin{equation}
  \int\mathrm{d}\lambda x(\lambda) f_{\alpha}(\lambda) \approx
  c_{\alpha} X.
\end{equation}
where $X$ is the $\alpha$-independent scaling function defined as 
\begin{equation}\label{eq:X}
  X=\int \mathrm{d}\lambda x(\lambda) \lambda^a.
\end{equation}
Here ${a>0}$ is the leading order of $f_{\alpha}(\lambda)$ for $\lambda$ close to $\lambda^{\ast}$, i.e.\ ${f_{\alpha}(\lambda)\propto (\lambda-\lambda^{\ast})^{a}}$. Since $X\to0$ in the limit $t\to\infty$, it plays the role of the small parameter for large times. By expanding the exponential in $X$, we observe that at leading order the coefficients $J_k$ have the following scaling 
\be\label{eq:J0k}
J_{0} = 1 +d_0 X, \qquad  J_{k\neq0} = d_k X,\qquad d_k\in\mathbb C\,. 
\ee
Inspecting the form of \eqref{eq:AsymmetryGeneral1} this means that the leading order contribution is obtained by {neglecting} the $k=0$ contribution, i.e. $
 \Delta S_A=
  - \sum_{k\neq 0} \Re[J_k \log J_k]$. Next, noting that, at leading order, each $d_k$ inside the logarithm can be replaced with $d_0$, we can perform the sum over $k$ by using $\sum_k J_k=1$. This yields the convenient expression  
\begin{equation}\label{eq:asymmetry_large_t}
  \Delta S_A= - (1-J_0) \log (1-J_0) + \mathcal O(X),
\end{equation}
where we used that $J_0$ is real because $f_\alpha^*(\lambda)=f_{-\alpha}(\lambda)$.
Finally we observe that, since $f(x)=-x\log x$ is monotonically increasing for small $x$, \ref{it:cond2} is equivalent to requiring $J_{0,2}<J_{0,1}$.

\section{Expansion of charged moments around $n=1$}
\label{sec:ExpansionInN}

Here we carry out an expansion of $\tr[\rhoAQ^n]/\tr[\rho_{A}^n]$ around $n=1$ and derive Eq.~\eqref{eq:charged_1} of the main text. We consider separately three cases: (i) free fermionic systems; (ii) Rule 54; (iii) generic integrable models. Even though the latter case includes the two the former ones, it is useful to consider them first as they involve more explicit expressions.

\subsection{Free Fermions}

In the case of free fermions, the expansion around $n=1$ of Eq.~\eqref{eq:charged_1} is exact. Let us see this for the particular quench protocol considered in the main text, taking as initial configuration the squeezed states 
\be
\ket{\Psi_0}={\rm exp}\left[-\int_0^{\pi} d\lambda \mathcal{M}(\lambda)\eta_\lambda^\dagger \eta_{-\lambda}^\dagger\right]\ket{0}
\ee
and as post-quench Hamiltonian  
\be
H=\int_{-\pi}^\pi d\lambda \epsilon(\lambda) \eta_\lambda^\dagger \eta_\lambda.
\ee
Since the initial state is Gaussian and the Hamiltonian is a quadratic fermionic operator, all the information about the time evolution after the quench is encoded in the two-point correlation functions $\bra{\Psi(t)} c_j^\dagger c_{j'}\ket{\Psi(t)}$ and $\bra{\Psi(t)}c_j c_{j'}\ket{\Psi(t)}$, where the real space creation and annihilation operators $c_j^\dagger$, $c_j$ are related to $\eta_\lambda^\dagger$, $\eta_\lambda$ by the Fourier transformation
\begin{equation}
 \eta_k=\sum_{j\in\mathbb{Z}} e^{ijk} c_j.
\end{equation}
In particular, in order to compute the Rényi entanglement asymmetry (cf. Eq.~\eqref{eq:Renyi_asymm}), one can consider the restriction of these correlators to the subsystem $A$,
\begin{equation}
\Gamma_{jj'}(t)=2\bra{\Psi(t)} \boldsymbol{c}_j^\dagger \boldsymbol{c}_{j'}\ket{\Psi(t)}-\delta_{jj'}, \quad j, j'\in A,
\end{equation}
where $\boldsymbol{c}_j=(c_j^\dagger, c_j)$. Then, as shown in detail in Ref.~\cite{ares2023lack}, the charged moments $Z_n(\boldsymbol{\alpha}, t)$ are given in terms of $\Gamma_{jj'}$ by 
\begin{equation}\label{eq:numerics}
Z_n({\boldsymbol{\alpha}, t})=\sqrt{\det\left[\left(\frac{I-\Gamma(t)}{2}\right)^n\left(I+\prod_{j=1}^n W_j(t)\right)\right]},
\end{equation}
with $W_j(t)=(I+\Gamma(t))(I-\Gamma(t))^{-1}e^{i\alpha_{jj+1}n_A}$ and $n_A$ is the diagonal matrix with non-zero entries $(n_A)_{2j, 2j}=1$ and $(n_A)_{2j-1, 2j-1}=-1$, $j=1, \dots, \ell$. In the quench studied in this work, $\Gamma_{jj'}$ is the $2\ell\times 2\ell$ block Toeplitz matrix
\begin{equation}
\Gamma_{jj'}(t)=\int_{-\pi}^\pi \frac{d\lambda}{2\pi}\mathcal{G}(\lambda, t) e^{-i\lambda(j-j')},
\end{equation}
where $\mathcal{G}(\lambda, t)$ is the $2\times 2$ matrix
\begin{equation}\label{eq:squeezes_symbol}
\mathcal{G}(\lambda, t)=\left(\begin{array}{cc}
1-2\vartheta(\lambda) & -2 e^{-i2t\epsilon(\lambda)}p(\lambda) \\
2e^{i2t\epsilon(\lambda)}p(\lambda) & 
2\vartheta(\lambda)-1
\end{array}
\right),
\end{equation}
with $\vartheta(\lambda)$ the density of excitations with momentum $\lambda$, and $p(\lambda)=\bra{\Psi_0}\eta_\lambda^\dagger \eta_{\pi-\lambda}^\dagger\ket{\Psi_0}=i \mathcal{M}(\lambda)/(1+\mathcal{M}(\lambda)^2)$.

Applying the results found in Ref.~\cite{ares2023lack} for a determinant that involves a product of block Toeplitz matrices as in Eq.~\eqref{eq:numerics}, we obtain that
at time $t=0$ and large subsystem size $\ell$, the charged moments 
behave as
\begin{equation}
Z_n(\boldsymbol{\alpha}, t=0)=Z_n(\boldsymbol{0}, t=0) e^{A_n(\boldsymbol{\alpha})\ell}.
\end{equation}
The coefficient $A_n(\boldsymbol{\alpha})$ factorizes in the replica space 
\begin{equation}
A_n(\boldsymbol{\alpha})=\sum_{j=1}^n\int_{-\pi}^\pi d\lambda  f_{\alpha_{j,j+1}}(\lambda),
\end{equation}
where $f_{\alpha}(\lambda)=\log[1-\vartheta(\lambda)+\vartheta(\lambda)e^{2i\alpha}]/(4\pi)$, and $\alpha_{j,j+1}=\alpha_j-\alpha_{j+1}$, with $\alpha_{n+1}\equiv \alpha_1$.

\begin{figure}[t!]
    \centering
     \includegraphics[width=0.45\textwidth]{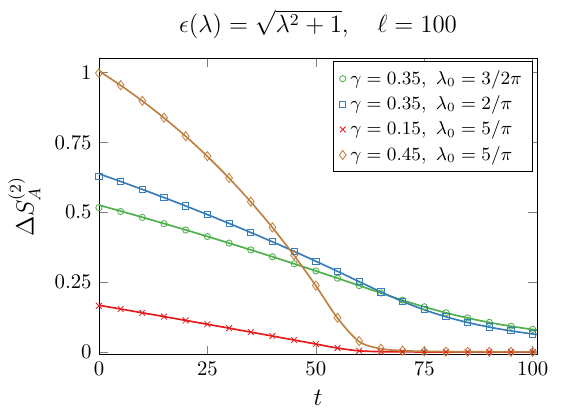}
     \includegraphics[width=0.45\textwidth]{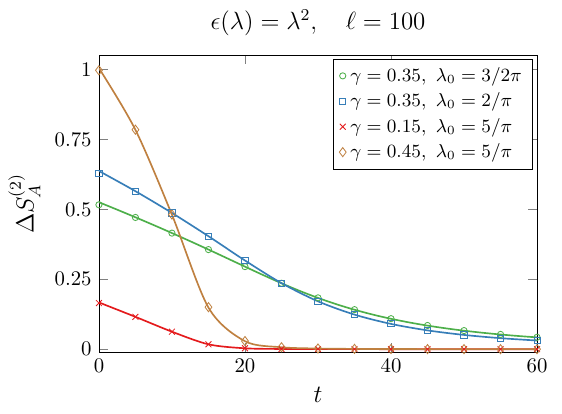}

     \caption{Time evolution of the R\'enyi entanglement asymmetry $\Delta S_{A}^{(2)}(t)$ after the quench from a squeezed state with $\mathcal{M}(\lambda)$ given in the main text and taking a post-quench free fermionic Hamiltonian with dispersion relation $\epsilon(\lambda)=\sqrt{\lambda^2+1}$ (left panel) and $\epsilon(\lambda)=\lambda^2$ (right panel). 
     The symbols are the exact numerical value of the asymmetry calculated using Eq.~\eqref{eq:numerics}. 
     The continuous lines are our prediction obtained using the analytic expression for the charged moments $Z_n(\boldsymbol{\alpha})$ reported in Eq. \eqref{eq:charged_moments_app}.}
     \label{fig:asymmetry_squeezed}
\end{figure}

Following Refs.~\cite{ares2022entanglement, ares2023lack}, we can  apply the quasiparticle picture of entanglement and the fact that the symmetry is restored in the limit $t\to\infty$ to deduce the full time evolution of the charged moments $Z_n(\boldsymbol{\alpha}, t)$ in the ballistic regime $t,\ell\to\infty$ with $\zeta=t/\ell$ finite. We find
\begin{equation}\label{eq:charged_moments_app}
Z_n(\boldsymbol{\alpha}, t)=Z_n(\boldsymbol{0}, t) e^{\ell(A_n(\boldsymbol{\alpha})+B_n(\boldsymbol{\alpha}, \zeta))},
\end{equation}
where
\begin{equation}
B_n(\boldsymbol{\alpha}, \zeta)=\sum_{j=1}^n \int_{-\pi}^\pi d\lambda\min(2\zeta |v(\lambda)|, 1)f_{\alpha_{j, j+1}}(\lambda),
\end{equation}
and $v(\lambda)$ is the group velocity of the excitations. By doing a change of variables, $\alpha_{j,j+1}\mapsto \alpha_j'$, with $\alpha_n'=-\sum_{j=1}^{n-1}\alpha_j'$, we can rewrite the charged moments above as
\begin{equation}
Z_n(\boldsymbol{\alpha'}, t)=Z_n(\boldsymbol{0}, t) \exp\left[\ell\sum_{j=1}^n \int_{-\pi}^\pi d\lambda x_\zeta(\lambda) f_{\alpha_{j}'}(\lambda)\right].
\end{equation}
Plugging it into Eq. \eqref{eq:FT}, we find
\begin{equation}
     I_n=\int_{-\pi}^{\pi}
    \frac{{\rm d}\boldsymbol{\alpha}}{(2\pi)^{n-1}} 
    \delta\bigl({\textstyle \sum_{j}}\alpha_j\bigr) \exp[ \ell \sum_{j=1}^{n} \int \!\!\mathrm{d}\lambda\, x_\zeta(\lambda)f_{\alpha_{j}}(\lambda)]\!,
    \label{eq:Inapp}
\end{equation}
which coincides with Eq. \eqref{eq:In} of the main text. Therefore, in this case, Eq.~\eqref{eq:charged_1} is exact $\forall n \in \mathbb{N}$.

In Fig.~\ref{fig:asymmetry_squeezed}, we check numerically the analytic expression of Eq.~\eqref{eq:charged_moments_app} for the charged moments. The symbols are the exact value of the $n=2$ Rényi entanglement asymmetry as a function of time computed using Eq.~\eqref{eq:numerics} for different quenches. The solid lines have been obtained using the prediction of Eq.~\eqref{eq:charged_moments_app} for $Z_n(\boldsymbol{\alpha}, t)$.

\subsection{Rule 54}

Let us now consider the quantum cellular automaton Rule 54~\cite{bobenko1993two, buca2021rule}, which has recently emerged as a very useful example of interacting integrable model leading to explicit exact results. This cellular automaton can be thought of as a quantum circuit of qubits (${d=2}$) with time evolution implemented by the unitary operator 
\begin{equation}
\hat{\mathbb{U}}=\hat{\Pi}^{\dagger} \hat{\mathbb{U}}_{\mathrm{e}} 
  \hat{\Pi} \hat{\mathbb{U}}_{\mathrm{e}},\qquad
\hat{\mathbb{U}}_{\mathrm{e}}= \prod_{j\in \mathbb{Z}} \hat{U}_{2j}.
\end{equation}
The ``local gate'' $\hat{U}_j$ acts non-trivially only on three sites ($j-1$, $j$, and $j+1$), implementing a deterministic update of the middle one
\begin{equation}
\mel{s_1^{\prime} s_2^{\prime} s_3^{\prime}}{\hat{U}}{s_1^{\phantom{\prime}} s_2^{\phantom{\prime}} s_3^{\phantom{\prime}}}
  =
  \delta_{s_1^{{\prime}},s_1}
  \delta_{s_2^{{\prime}},\chi(s_1,s_2,s_3)}
  \delta_{s_3^{{\prime}},s_3},\qquad
  \chi(s_1,s_2,s_3)\equiv s_1+s_2+s_3+s_1 s_3\pmod{2}.
\end{equation}
Considering a quench from a \emph{solvable} initial state~\cite{klobas2021exactrelaxation} 
\be
\ket*{\Psi_0}=(\sqrt{1-\vartheta}\ket{00}+\sqrt{\vartheta}\ket{01})^{\otimes L/2},
\ee
where $\vartheta\in[0,1]$ determines the quasiparticle occupations, one can characterise exactly the charged moments of
\be
Q=\sum_j  \frac{1-\sigma_{j+1}^z}{4}
  \left(2+\sigma_{j+2}^z+\sigma_j^z\sigma_{j+2}^z\right), 
\ee
in two regimes: (i) $t \leq \ell/2$ and (ii) $t\geq 3 \ell/2$~\cite{klobas2023inprep}. In the regime (i) we find
\begin{equation}
  Z_{n}(\bm{\alpha},t)\approx Z_{n}(\bm{0},t)\,
  \Lambda_{\bm{\alpha}}^{\ell-2t}
  \left(\frac{\lambda_{\bm{\alpha}}}{\lambda_{\bm{0}}}\right)^{2t},\qquad
  \Lambda_{\bm{\alpha}}=\prod_{j=1}^{n} (1-\vartheta+\vartheta e^{2i\alpha_{j}}), \qquad \sum_j \alpha_j=0\,,
\end{equation}
where $\approx$ denotes the leading order in $t$ and $\ell$, and $\lambda_{\bm{\alpha}}$ is the largest-magnitude solution to the cubic equation
\begin{equation}\label{eq:r54cubicEq}
  x^3=\left(x (1-\vartheta)^n+\vartheta^n \Lambda_{\bm{\alpha}}\right)^2.
\end{equation}
On the other hand, in the regime (ii) we obtain 
\be
   Z_{n}(\bm{\alpha},\zeta l) \approx d_n^{2\ell}, 
\ee
where we set 
\be
  d_n=(1-\vartheta)^n+\vartheta^n.
\ee
Therefore, in this case we can write \eqref{eq:charged_1} of the main text as 
\begin{equation} \label{eq:r54expression}
  \frac{\tr(\rhoAQ^n)}{\tr(\rho_{A}^n)}=
  \int_{-\pi}^{\pi}\frac{\mathrm{d}\alpha_1}{2\pi}
  \int_{-\pi}^{\pi}\frac{\mathrm{d}\alpha_2}{2\pi}
  \cdots
  \int_{-\pi}^{\pi}\frac{\mathrm{d}\alpha_n}{2\pi}
  \delta_p\bigl({\textstyle \sum_{j}}\alpha_j\bigr)
  e^{\ell B_n(\bm{\alpha},\zeta)}.
\end{equation}
where we set $\zeta={t}/{\ell}$ and 
\begin{equation}
  B_n(\bm{\alpha},\zeta) = \frac{1}{\ell}\log \left(\frac{Z_{n}(\bm{\alpha},t)}{Z_{n}(\bm{0},t)}\right) = \begin{cases}
  (1-2\zeta) \log \Lambda_{\bm{\alpha}}
  +2\zeta \log\frac{\lambda_{\bm{\alpha}}}{\lambda_{\bm{0}}}, & \zeta \leq 1/2\\
  \\
 0 & \zeta\geq 3/2 \\
  \end{cases}.
\end{equation}

To find the behaviour of asymmetry in the $n\to 1$ limit we expand
$B_n(\bm{\alpha},\zeta)$ up to the first order in $(n-1)$, i.e.
\begin{equation}\label{eq:firstorderRule54Everything}
  \lim_{n\to 1}
  \frac{\tr(\rhoAQ^n)}{\tr(\rho_A^n)}=
  \lim_{n\to 1}
  \int_{-\pi}^{\pi}\frac{\mathrm{d}\alpha_1}{2\pi}
  \cdots
  \int_{-\pi}^{\pi}\frac{\mathrm{d}\alpha_{n-1}}{2\pi}
  e^{\ell  (B_1(0,\zeta)+(n-1)\left[\partial_{n}B_n(\bm{\alpha},\zeta)\right]_{n\to 1})}.
\end{equation}
To evaluate the above expression we first note that, by definition, the $n\to 1$ limit of $\Lambda_{\bm{\alpha}}$
and $\lambda_{\bm{\alpha}}$ is independent of the value of $\bm{\alpha}$ and is
given by
\begin{equation}
  \lim_{n\to 1}\Lambda_{\bm{\alpha}}=\lim_{n\to 1}\Lambda_{\bm{0}}=1,\qquad
  \lim_{n\to 1}\lambda_{\bm{\alpha}}=\lim_{n\to 1}\lambda_{\bm{0}}=1,
\end{equation}
which gives 
\begin{equation}
  \lim_{n\to 1} B_n(\bm{\alpha},\zeta) = B_1(0,\zeta)= 0,
\end{equation}
for all $\zeta$. The next step is to evaluate $\partial_n B_n$ for $n=1$. Considering the early time regime $\zeta<\frac{1}{2}$ we find 
\begin{equation}
  \left[\partial_{n} B_n(\bm{\alpha},\zeta)\right]_{n\to 1}=
  \lim_{n\to 1}
  \left[
    \frac{\Lambda^{\prime}_{\bm{\alpha}}}{\Lambda_{\bm{\alpha}}}
    +2\zeta \frac{\lambda_{\bm{\alpha}}^{\prime}}{\lambda_{\bm{\alpha}}}
    -2\zeta \frac{\lambda_{\bm{0}}^{\prime}}{\lambda_{\bm{0}}}
    \right],
\end{equation}
where $\cdot^{\prime}$ here denotes the derivative with respect to the R\'enyi
index $n$. At first sight it is not obvious how to understand these terms.
However, assuming that $\Lambda_{\bm{\alpha}}^{\prime}$ can be meaningfully
defined, we can express $\lambda_{\bm{\alpha}}^{\prime}$ in terms of it and we
obtain
\begin{equation}\label{eq:r54EarlyTimeRegimento1B}
  \left[\partial_{n} B_n(\bm{\alpha},\zeta)\right]_{n\to 1}=
  \left(1-\frac{2\zeta}{1+2\vartheta}\right)
  \left(\lim_{n\to 1}\Lambda_{\bm{\alpha}}^{\prime}\right)=
  \left(1-2\zeta v\right) f(\bm{\alpha}),
\end{equation}
where we took into account that $(1+2\vartheta)^{-1}$ is the expression for the
dressed velocity $v$ of quasiparticles in the stationary
state~\cite{gopalakrishnan2018hydrodynamics}, and we introduced
$f(\bm{\alpha})$ for the $n\to 1$ limit of $\Lambda_{\bm{\alpha}}^{\prime}$. To find its value, we first
expand $\log\Lambda_{\bm{\alpha}}$ around $n=1$ up to the first order,
\begin{equation}
  \begin{aligned}
    \log\Lambda_{\bm{\alpha}}&=
  \lim_{n\to 1} \log\Lambda_{\bm{\alpha}}
  +(n-1) \lim_{n\to 1} \left(\log\Lambda_{\bm{\alpha}}\right)^{\prime}
    +\mathcal{O}((n-1)^2) \\ &= 
  \lim_{n\to 1} \log\Lambda_{\bm{\alpha}}
  +(n-1) \lim_{n\to 1} \frac{\Lambda_{\bm{\alpha}}^{\prime}}{\Lambda_{\bm{\alpha}}}
    +\mathcal{O}((n-1)^2) \\&= 
  (n-1) f(\bm{\alpha}) +\mathcal{O}((n-1)^2),
  \end{aligned}
\end{equation}
where the second to last equality follows from $\lim_{n\to 1}\Lambda_{\bm{\alpha}}=1$, and
the definition of $f(\bm{\alpha})$. Therefore, putting all together we find 
\be
B_1(0,\zeta)+(n-1)\left[\partial_{n}B_n(\bm{\alpha},\zeta)\right]_{n\to 1}  = \left(1-2\zeta v\right) \log\Lambda_{\bm{\alpha}}
\ee
On the other hand, this quantity is clearly 0 for $\zeta\geq 3/2$. Continuing by continuity to $1/2\leq \zeta \leq 3/2$ by invoking the quasiparticle picture we obtain the quasiparticle looking expression  
\be
B_1(0,\zeta)+(n-1)\left[\partial_{n}B_n(\bm{\alpha},\zeta)\right]_{n\to 1}  = {\rm max}\left(1-2\zeta v,0\right) \log\Lambda_{\bm{\alpha}}. 
\ee 
This finally gives us
\begin{equation}
  \frac{\tr[\rhoAQ^n]}{\tr[\rho_A^n]}=I_n(\mathrm{max}\{\ell-2vt,0\})+\mathcal{O}((n-1)^2),
  \qquad
  I_n(z)=
  \int_{-\pi}^{\pi}\frac{\mathrm{d}\alpha_1}{2\pi}
  \cdots
  \int_{-\pi}^{\pi}\frac{\mathrm{d}\alpha_{n}}{2\pi} 2\pi \delta_p\bigl({\textstyle \sum_{j}}\alpha_j\bigr)
  e^{z \log\Lambda_{\bm{\alpha}}}.
\end{equation}
Namely, we obtain Eq.~\eqref{eq:charged_1} of the main text with 
\be
f_{\alpha,\nu} = \frac{1}{2}\log\Lambda_{{\alpha}}= \frac{1}{2} \log (1-\vartheta+\vartheta e^{2i\alpha})\,. 
\ee
It is immediate to verify that $f^*_{\alpha,\nu}=f^{\phantom{\ast}}_{-\alpha,\nu}$ and $f_{\alpha,\nu} = f_{\alpha+\pi,\nu}$.

\subsection{General Interacting Integrable Systems}
In the preceeding sections we have seen how to obtain the leading behaviour of the charged moments in the limit $n\to 1$ for two different models using distinct methods.  For a generic integrable model we cannot utilize either of these methods as they are specialized to their particular cases and so we must resort to a third method which can nevertheless reproduce the results of those sections but is less rigorous.

We recall that an explicit expression for the charged moments in integrable models at short times $t\ll \ell$ has been derived in~\cite{bertini2023dynamics} while the behavior of $\tr[\rho_A^n(t)]$ was found in~\cite{bertini2022growth}.  Our strategy will be to use these expressions  for $\alpha_j\in i \mathbb{R}$ and perform an expansion about $\alpha_j\approx 0$.  From this we can formulate the quasiparticle picture to capture the full time dynamics, then resum the series and analytically continue it back to $\alpha_j\in \mathbb{R}$.  The validity of this last step can be questioned as we do  not have full analytic control over all terms in the expansion however we justify it by comparing to the exact results of the previous sections.

To begin we state some facts about integrable models. They exhibit a set of stable quasiparticle excitations specifed by a species index $m$ and a rapidity variable $\lambda$ (in the free model above $M=1$ and $\lambda\in\mathbb{R}$ whereas for Rule 54 we have $M=1$ and the rapidity is discrete $\lambda=\pm 1$).  A state of the system is described through a set of distributions for these quasiparticles, in particular $\vartheta_m(\lambda)\in [0,1]$ is the occupation function of quasparticle species $m$ at rapidity $\lambda$, $\rho^t_m(\lambda)$ is their density of states and $\rho_m(\lambda)=\vartheta_m(\lambda)\rho^t_m(\lambda)$ is the distribution of occupied modes.  These quasiparticles scatter nontrivially through the scattering kernel $T_{ml}(\lambda,
\mu)$ (which is the log derivative of their scattering matrix) for species $m,l$ and rapidities $\lambda,\mu$, typically  $T_{ml}(\lambda,
mu)=T_{ml}(\lambda-\mu)$.  Each quasiparticle carries a bare charge $q_m$ under the $U(1)$ symmetry.  This is dressed by the interactions after  which it is denoted $q_{{\rm eff}, m}(\lambda)$. Similarly each quasiparticle has a bare velocity which is dressed also, we denote this by $v_m(\lambda)$. 

We rotate $\alpha_j\to -i\alpha_j$ ($\alpha_n=-\sum_{j=1}^{n-1}\alpha_j$) and write the initial value of the charged moments for a model with $M$ species as
\begin{eqnarray}
\frac{1}{\ell}\log  \tr[\rhoAQ^n(0)]&=& \frac{1}{2}\sum_{j=1}^{n}\sum_{m=1}^M\int {\rm d}\lambda\, \rho^t_m(\lambda)\left[\mathcal{K}^{2\alpha_j}_{1,m}(\lambda)+\vartheta_{m}(\lambda)\log w_{1,m}^{2\alpha_j}e^{2\alpha_j}\right]\\
&\equiv &\frac{1}{2}\sum_{j=1}^{n}\sum_{m=1}^M\int {\rm d}\lambda\,d^{2\alpha_j}_{1,m}(\lambda)\\
\mathcal{K}^{\beta}_{n,m}(\lambda)&=&\log\left[(1-\vartheta_m(\lambda))^n+\vartheta_m^n(\lambda)e^{-\log w^{\beta}_{n,m}}\right],\\
\log w^{\beta}_{n,m}(\lambda)&=&-\beta q_m+\int{\rm d}\mu T_{ml}(\lambda-\mu)\mathcal{K}_{n,l}^{\beta}(\mu)
\end{eqnarray}
At long times since the symmetry is restored the charged moment is also straightforward to evaluate and we have 
\begin{eqnarray}
\lim_{t\to\infty}\frac{1}{\ell}\log  \tr[\rhoAQ^n(0)]&=&\sum_{j=1}^{n}\sum_{m=1}^M\int {\rm d}\lambda\, \rho^t_{n,m}(\lambda)\left[\mathcal{K}^{0}_{n,m}(\lambda)+\vartheta_{n,m}(\lambda)\log w_{n,m}^{0}\right]\\
&\equiv &\sum_{j=1}^{n}\sum_{m=1}^M\int {\rm d}\lambda\,d^{\,0}_{n,m}(\lambda)\\
\end{eqnarray}
where 
\begin{eqnarray}
\vartheta_{n,m}(\lambda)=\frac{(\vartheta_m)^n}{(\vartheta_m)^n+(1-\vartheta_m)^ne^{\log w^{\, 0}_{n,m}}}
\end{eqnarray}
and $\rho^t_{n,m}$ is the associated density of states. 

At time scales $t\ll \ell$ we have that the charged moments behave as 
\begin{eqnarray}
\frac{1}{t}\log  \frac{\tr[\rhoAQ^n(t)]}{\tr[\rhoAQ^n(0)]}&=&\sum_{m=1}^M\int \mathrm{d}\lambda \,s^{\boldsymbol \alpha}_{m}(\lambda)
\end{eqnarray}
where $s^{\boldsymbol \alpha}_{m}(\lambda)$ is a complicated but known function expressible in terms of a set of TBA like equations and the distributions $\vartheta_m(\lambda)$.  Without specifying it however we can form the quasiparticle picture for the full time dynamics.  In  particular the full time dynamics is given by
\begin{eqnarray}
\log \tr[\rhoAQ^n(t)]=\frac{1}{2}\ell\sum_{j=1}^n\sum_{m=1}^M\int {\rm d}\lambda\, d^{2\alpha_j}_{1,m}(\lambda)+\sum_{m=1}^M\int{\rm d}\lambda \,{\rm min}[ t s^{\boldsymbol \alpha}_{m}(\lambda),\ell(d^{\,0}_{n,m}(\lambda)-\frac{1}{2}\sum_{j}^n d^{2\alpha_j}_{1,m}(\lambda))]
\end{eqnarray}
which can be readily checked to reproduce the formuale above in the appropriate limits.  

At this point we note that the explicit expression for $s^{\boldsymbol{\alpha}}_m(\lambda)$ given in~\cite{bertini2023dynamics} has the following properties 
\begin{eqnarray}
\partial_{\alpha_j}s^{\boldsymbol{\alpha}}_m(\lambda)|_{\boldsymbol{\alpha}=0}=2 |v_m(\lambda)|\partial_{\alpha_j}\sum_j^n d^{2\alpha_j}_{1,m}(\lambda)|_{\boldsymbol{\alpha}=0}\\
\partial_{\alpha_j}\partial_{\alpha_k}s^{\boldsymbol{\alpha}}_m(\lambda)|_{\boldsymbol{\alpha}=0}=2  |v_m(\lambda)|\partial_{\alpha_j}\partial_{\alpha_k}\sum_j^n d^{2\alpha_j}_{1,m}(\lambda)|_{\boldsymbol{\alpha}=0}
\end{eqnarray}
and similarly for the higher derivatives. This means that in the region of $\alpha_j=0$ the complicated expression for $s^{\boldsymbol \alpha}_{m}(\lambda)$ can be replaced by $ 2 |v_m(\lambda)|\sum_j^n d^{2\alpha_j}_{1,m}(\lambda)+s^{\boldsymbol 0}_{m}(\lambda)$ which is easier to deal with.  Combining these together we find that the desired expression is 
\begin{eqnarray}
\log\frac{ \tr[\rhoAQ^n(t)]}{ \tr[\rho_A^n(t)]}&=&\ell\sum_{j=1}^n\sum_{m=1}^M\int \mathrm{d}\lambda\, x_{m,\zeta}(\lambda)\left[\mathcal{K}^{2\alpha_j}_{1,m}(\lambda)+\vartheta_{m}(\lambda)\log w_{1,m}^{2\alpha_j}e^{2\alpha_j}\right],\\
x_{m,\zeta}(\lambda)&=&{\rm max}[1-2 \zeta v_m(\lambda),0]
\end{eqnarray}
This can then be safely rotated back to $\alpha_j\to i\alpha_j$ to obtain the charged moments in the region of $n=1$. Specializing to the case of $M=1$ discussed in the main text we have that 
\begin{eqnarray}
f_{\alpha}=\rho^t(\lambda)\left(\log{\left[1-\vartheta(\lambda)+\vartheta(\lambda)e^{-\log w^{2i\alpha}(\lambda)}\right]}+\vartheta(\lambda)\log w^{2i\alpha}(\lambda)e^{2i\alpha}\right)/2.
\end{eqnarray}
This can be checked to reproduce the earlier results in the free case and for Rule 54. 

\section{Free fermions: Large time asymptotic expressions for $\Delta S^{(n)}_A$}
\label{sec:FFgenericn}

The goal of this Appendix is to study the large time behaviour of the 
entanglement asymmetry $\Delta S^{(n)}_A(t)$ in the free fermionic case for a generic Rényi index $n\in \mathbb{N}$ and check that the analytic continuation of this result agrees with the general replica limit $n\to 1$ in Eq. \eqref{eq:asymmetry_large_t}. 

For free fermions, the function $f_\alpha(\lambda)$ in 
\begin{equation}
  J_k\!=\!\!\int_{-\pi}^{\pi}\frac{\mathrm{d}\alpha}{2\pi}
    e^{i k \alpha}
  \exp[\ell\int \!\!\mathrm{d}\lambda\, x_\zeta(\lambda) f_{\alpha}(\lambda)]\!
\end{equation}
takes the form
\begin{equation}
f_\alpha(\lambda)=\frac{1}{4\pi}\log(1-\vartheta(\lambda)+\vartheta(\lambda)e^{2i\alpha}).
\end{equation}

In the regime $\zeta \to \infty$, due to the presence of the function $x_\zeta(\lambda)$ in the integral above, the argument of the exponential becomes small and we can use its series expansion to get
\begin{equation}\label{eq:inf_order2}
\begin{aligned}
 J_k&=
  \int_{-\pi}^{\pi}\frac{d\alpha
  }{2\pi}e^{ik\alpha}\sum_{r=0}^{\infty}\frac{\ell^r}{r!} \Big[\int_{-\pi}^{\pi}\frac{d\lambda_1}{4\pi}x_\zeta(\lambda_1)\log\left[ (1-\vartheta(\lambda_1)+\vartheta(\lambda_1)e^{2i\alpha}\right]\ldots\\
&\times \int_{-\pi}^{\pi}\frac{d\lambda_r}{4\pi}x_\zeta(\lambda_r)\log\left[ (1-\vartheta(\lambda_r)+\vartheta(\lambda_r)e^{2i\alpha})\right]\Big].
\end{aligned}
\end{equation}

Since we are interested in the asymptotic expression of $\Delta S_A^{(n)}$ at large $\zeta$, we can restrict the sum above to the term $O(\ell)$,
\begin{equation}
J_k\approx \delta_{k, 0}+\ell\int_{-\pi}^\pi \frac{d\alpha}{2\pi}
e^{i\alpha k}\int_{-\pi}^\pi\frac{d\lambda}{4\pi}x_\zeta(\lambda)\log(1-\vartheta(\lambda)+\vartheta(\lambda)e^{2i\alpha}).
\end{equation}
We can now use the series expansion of the logarithm function to rewrite the argument of the integral as
\begin{equation}\label{eq:binom}
\log(1-\vartheta(\lambda)+\vartheta(\lambda)e^{2i\alpha})= \mathrm{max}[ \log(\vartheta(\lambda)),\log(1-\vartheta(\lambda))]-\sum_{s=1}^{\infty}\frac{\mathrm{min}[ \vartheta(\lambda),1-\vartheta(\lambda)]^s}{s \mathrm{max}[ \vartheta(\lambda),1-\vartheta(\lambda)]^s}(-1)^se^{2i\alpha s}.
\end{equation}
Then, taking into account that $s$ in Eq. \eqref{eq:binom} can only assume positive integer values, we find that $J_{2k-1}=0$ while for $k$ even we have at $O(\ell)$
\begin{multline}\label{eq:J2k_asymptotics}
  J_{2k}\approx\delta_{k,0}+\ell\int_{-\pi}^{\pi}\frac{d\lambda}{4\pi}x_\zeta(\lambda)\mathrm{max}[ \log(\vartheta(\lambda)),\log(1-\vartheta(\lambda))]\delta_{k,0}\\
  +(1-\delta_{k,0})(-1)^k\Theta(-k)\frac{\ell}{k}\int_{-\pi}^{\pi}\frac{d\lambda}{4\pi}x_\zeta(\lambda)\left(\frac{\mathrm{max}[ \vartheta(\lambda),1-\vartheta(\lambda)]}{\mathrm{min}[ \vartheta(\lambda),1-\vartheta(\lambda)]}\right)^k,
\end{multline}
where $\Theta(x)$ is the Heaviside theta function. This analysis shows that the functions $J_k$ are real and, therefore, Eq. \eqref{eq:defJk} admits a unique analytic continuation given by
\begin{equation}\label{eq:anal_cont_app}
    I_n\to I_z, \qquad I_z=\sum_{k=-\infty}^{0}J^z_{2k}, \qquad z \in \mathbb{C}.
\end{equation}
The fact that $J_{k}$ is a real function guarantees that the hypothesis of the Carlson theorem \cite{carlson} are satisfied and that the analytic continuation we have chosen in Eq. \eqref{eq:analyticcont} is the unique and correct one in the case of free fermions. This also gives an independent check of the general result in Eq. \eqref{eq:asymmetry_large_t}, which has been derived assuming the validity of Eq. \eqref{eq:analyticcont}. Moreover, observe that the terms $J_{2k}$ in Eq. \eqref{eq:J2k_asymptotics} are of the form in Eq. \eqref{eq:J0k} but now we have explicitly proven that, for free fermions, the coefficients $d_k$ are real numbers. 

Before concluding this Appendix, we show that the result in Eq. \eqref{eq:J2k_asymptotics} also allows us to obtain the asymptotic expression of $\Delta S_A^{(n)}$ for integer $n\geq 2$. Indeed, at leading order, we find 
\begin{equation}
    I_n=\sum_{k=-\infty}^0J_{2k}^n\simeq 1+n\ell \int_{-\pi}^{\pi}\frac{d\lambda}{4\pi}x_\zeta(\lambda)\mathrm{max}[ \log(\vartheta(\lambda)),\log(1-\vartheta(\lambda))],
\end{equation}
and, therefore, the Rényi entanglement asymmetry behaves at large times as
\begin{equation}\label{eq:asy_genericn}
    \Delta S_A^{(n)}(t)=\frac{n\ell}{1-n}\int_{-\pi}^{\pi}\frac{d\lambda}{4\pi}x_\zeta(\lambda)\mathrm{max}[ \log(\vartheta(\lambda)),\log(1-\vartheta(\lambda))].
\end{equation}

We can specify Eq. \eqref{eq:asy_genericn} in the setup of Ref. \cite{ares2022entanglement}, i.e. an infinite spin chain prepared in the tilted ferromagnetic state, where the angle $\theta$ controls the $U(1)$ symmetry breaking, which is dynamical restored after a quench to the XX spin chain. 
In this case, the occupation number $ \vartheta(\lambda)$ reads
\begin{eqnarray}
    \vartheta(\lambda)=\frac{1-\cos\Delta_\lambda}{2}, \qquad \cos \Delta_\lambda=\frac{2\cos(\theta)-(1+\cos^2\theta)\cos(\lambda)}{1+\cos^2\theta-2\cos(\theta)\cos(\lambda)}.
\end{eqnarray}
Therefore, Eq. \eqref{eq:asy_genericn} can be explicitly evaluated and we get
\begin{equation}
    \Delta S_{A}^{(n)}(t)\simeq \frac{n b(\zeta)\ell}{8(n-1)},\qquad b(\zeta)=\int_{-\pi}^\pi \frac{{\rm d} \lambda}{2\pi}x_\zeta(\lambda)\sin^2\Delta_\lambda.\end{equation}
From this result, we conclude that the prefactor $\pi^2/24$ in Eq. (13) of Ref. \cite{ares2022entanglement} is not correct and should be $n/(8(n-1))$ and, as a consequence, the prefactor $\pi/(1152)$ in Eq. (14) should be instead $n/(384\pi(n-1))$. We stress that the dependence on the tilting angle $\theta$ is the same, but the calculation of the $n$-dependence of the Rényi entanglement asymmetry was incorrect.

\section{Details on Lieb-Liniger Quench}
\label{sec:LLdetails}
Here we present some details on the quench dynamics of the entanglement entropy for the Lieb Liniger model quenched from the coherent BEC state,
\begin{eqnarray}
\ket{\Psi_0}&=&e^{-\frac{dL}{2}+\sqrt{d}b^\dag_0}\ket{0},~~b^\dag_0=\int_0^L \mathrm{d}x \, b^\dag(x),\\
&& d=\frac{1}{L}\int_0^L\mathrm{d}x\,\bra{\Psi_0}b^\dag(x)b(x)\ket{\Psi_0}.
\end{eqnarray}
The quench from the state $(b^\dag_0)^N\ket{0}$ was considered in~\cite{denardis2014solution} and many analytic formulae derived there can be applied in this case also.  In particular the long time post quench state is specified by the occupation function 
\begin{eqnarray}
\vartheta(\lambda)&=&\frac{a(\lambda)}{1+a(\lambda)},\\
a(\lambda)&=&\frac{2\pi d}{\lambda\sinh(2\pi\lambda/c)}I_{1-2i \lambda/c}(4\sqrt{d/c})I_{1+2i \lambda/c}(4\sqrt{d/c}).
\end{eqnarray}
From these one can also determine the rapidity distribution $\rho(\lambda)$, the density of states $\rho^t(\lambda)$ and the effective charge $q_{\text{eff},m}(\lambda)$
\begin{eqnarray}
q_{\text{eff}}(\lambda)&=&\frac{d}{2}\partial_d \log a(\lambda)=2\pi \rho^t(\lambda),\\\rho(\lambda)&=&\vartheta(\lambda)\rho^t(\lambda).
\end{eqnarray}
Using the results of SM~\ref{sec:ExpansionInN} for generic integrable models specified to the Lieb Liniger we find that the asymmetry is governed by 
\begin{eqnarray}
x_\zeta(\lambda)&=&\text{max}[1-2\zeta |v(\lambda)|,0]\\
f_{\alpha}(\lambda)&=&\rho^t(\lambda)\left(\log{\left[1-\vartheta(\lambda)+\vartheta(\lambda)e^{-\log w^{2 i\alpha}(\lambda)}\right]}+\vartheta(\lambda)\log w^{2 i\alpha}(\lambda)e^{2i\alpha}\right)/2
\end{eqnarray}
Here the function $w^{2 i\alpha}(\lambda)$ can be obtained by analytic continuation of $a(\lambda)$ to give
\begin{eqnarray}
w^{2 i\alpha}(\lambda)=e^{-i \alpha}\frac{I_{1-2i \lambda/c}(4\sqrt{d/c})I_{1+2i \lambda/c}(4\sqrt{d/c})}{I_{1-2i \lambda/c}(4\sqrt{de^{i\alpha}/c})I_{1+2i \lambda/c}(4\sqrt{de^{i\alpha}/c})}
\end{eqnarray}
which is valid for $|\alpha|<\pi/2$ and can be extended to $|\alpha|<\pi$  using the fact that $w^{2i\alpha}$ is invariant under $\alpha\to \alpha+\pi$.   We also require the dressed quasiparticle velocity $v(\lambda)$ which satisifies the following integral equation
\begin{eqnarray}
\rho^t(\lambda)v(\lambda)=\frac{\lambda}{\pi}+\frac{1}{\pi}\int {\rm d}\mu\,\frac{c}{(\lambda-\mu)^+c^2}\rho(\mu)v(\mu).
\end{eqnarray}
An analytic solution of this equation is not available but can nevertheless be evaluated numerically using standard techniques. 

\begin{figure}[t!]
    \centering
     \includegraphics[width=0.45\textwidth]{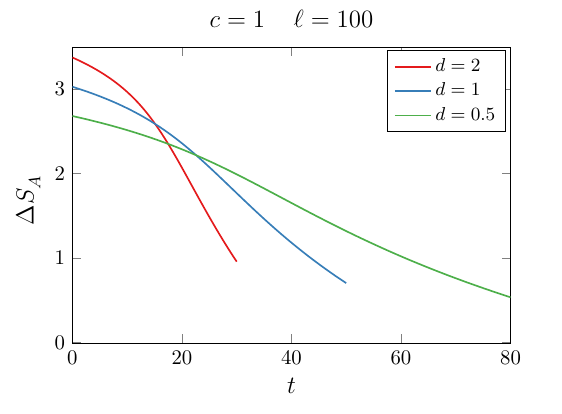}
     \includegraphics[width=0.45\textwidth]{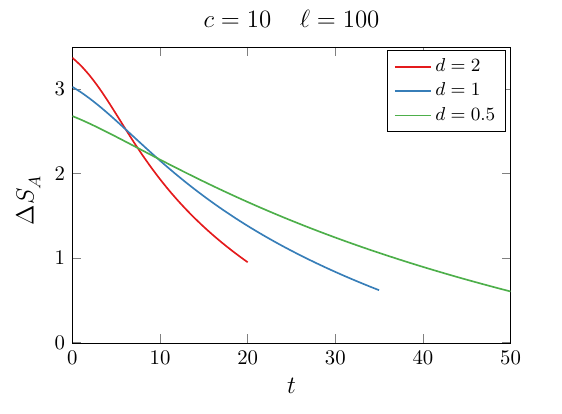}

     \caption{Entanglement asymmetry using the saddle point approximation for the Lieb Liniger model. On the left we quench from $d=0.5,1,2$ at $c=1$ and $\ell=100 $ while on the right we show the same at $c=10$. In both cases we see a crossing of the lines indicative of QME. We cut off the curves around $\Delta S_A(t)\sim \mathcal{O}(1)$ so as to stay within the regime of validity of the saddle point approximation. Note that the Mpemba times are longer for smaller $c$ and in fact diverge as $c\to 0$ when $\ket{\Psi_0}$ becomes an eigenstate of the model.  }
     \label{fig:asymmetry_lieb_liniger}
\end{figure}

The saddle point expression for the asymmetry can be readily found using these formulae and we obtain,
\begin{eqnarray}
\Delta S_A(t)=\frac{1}{2}+\frac{1}{2}\log{\pi\sigma_\zeta^2},\\
2\sigma^2_\zeta=\ell \int {\rm d}\lambda x_\zeta(\lambda)q_{\rm eff}^2(\lambda)\rho(\lambda)(1-\vartheta(\lambda))
\end{eqnarray}
At $t=0$ $x_0(\lambda)=1$ and using the anayltic expressions presented above  we find 
\begin{eqnarray}
\Delta S_A(0)=\frac{1}{2}+\frac{1}{2}\log{\pi \ell d/2}.
\end{eqnarray}
At finite time we can evaluate $\sigma^2_\zeta(\lambda)$ numerically and determine $\Delta S_A(t)$.  We plot the results in Fig.~\ref{fig:asymmetry_lieb_liniger} for  different values of $d$ and $c$. From this  we see that the QME always occurs between any two different values of $d$.  That is while  i.e.  $\Delta S[\rho_{A,1}(0)]>\Delta S[\rho_{A,2}(0)]$ for $d_1>d_2$ there is some time (within the validity of the saddle point approximation) after which  $\Delta S[\rho_{A,2}(t)]>\Delta S[\rho_{A,1}(t)]$. 

To confirm this extends beyond the saddle point regime and check condition (ii) is satisfied we study the long time behavior of $J_0$,
\begin{eqnarray}
J_0=\frac{1}{2\pi}\int_{-\pi}^{\pi} {\rm d}\alpha\, e^{\ell \int_{-\Lambda_\zeta} ^{\Lambda_\zeta} \mathrm{d}\lambda\,[1-\frac{|v(\lambda)|}{v(\Lambda_\zeta)}]f_{\alpha}(\lambda)}
\end{eqnarray}
where  $v(\Lambda_\zeta)=1/(2\zeta)$.  At long times $\Lambda_\zeta\to 0$ and we can expand the exponent to obtain
\begin{eqnarray}
J_0=1+\frac{\ell}{2\pi}\int_{-\pi}^{\pi} {\rm d}\alpha \int_{-\Lambda_\zeta} ^{\Lambda_\zeta} \mathrm{d}\lambda\,\left[1-\frac{|v(\lambda)|}{v(\Lambda_\zeta)}\right]f_{\alpha}(\lambda)
\end{eqnarray}
Furthermore expanding the integrand about $\lambda=0$ we have to leading order in $\Lambda_\zeta$
\begin{eqnarray}
J_0=1-\frac{\ell \Lambda_\zeta^3}{24}|\vartheta''(0)| \rho^t(0)[1+g(d/c)]
\end{eqnarray}
where here $\vartheta''(0)=\partial_\lambda^2\vartheta(\lambda)|_{\lambda=0}$, $\Lambda_\zeta=1/2\zeta v'(0)$ and $g(d)$ is 
\begin{eqnarray}
g(d)=\int_{-\pi}^\pi {\rm d}\alpha \,e^{i\alpha} \frac{I_1(4\sqrt{d e^{i \alpha}/c})^2}{I_1(4\sqrt{d /c})^2}-\log e^{-i \alpha}\frac{I_1(4\sqrt{d e^{i \alpha}/c})^2}{I_1(4\sqrt{d /c})^2}
\end{eqnarray}
We can examine this analytically in two limits, $d/c\ll 1$ and $d/c\gg 1$ in the first case only using the expansion of the Bessel function we find that 
\begin{eqnarray}
g(d/c)\approx 4 \frac{d}{c}+\mathcal{O}(d^2/c^2),~~d/c \ll 1
\end{eqnarray}
In the other case we find 
\begin{eqnarray}
g(d/c)\approx 8(1-2/\pi) \sqrt{\frac{d}{c}},~~d/c\gg 1
\end{eqnarray}
Now to evaluate $J_0$ in these limits we can note that
\begin{eqnarray}
\vartheta''(0)=-\frac{1}{d c(I_1(4\sqrt{d/c}))^2},~~~
\rho^t(0)=\frac{1}{4\pi}+\sqrt{\frac{d}{c}}\frac{I_0(4\sqrt{d/c})+I_2(4\sqrt{d/c})}{2 \pi I_1(4\sqrt{d/c})}
\end{eqnarray}
and also that $1 \leq v'(0)\leq 2$ with the upper and lower bounds achieved in the limits $d/c\to 0$ and $d/c\to\infty$.  From these considerations we can then determine that $1-J_0$ is a rapidly decreasing function function of $d/c$.  Moreover for the purposes of condition (ii) one can drop the $g(d)$ term and use 
\begin{eqnarray}
1-J_0\propto  \frac{\ell \Lambda_\zeta^3}{24}|\vartheta''(0)| \rho^t(0). 
\end{eqnarray}  
Therefore denoting by $J_{0,s},~s=1,2$ the value of $J_0$ for the states $1$ and $2$ with $d_1>d_2$ we have that
\begin{eqnarray}
J_{0,2}>J_{0,1}
\end{eqnarray}
which ensures that $\Delta S_{A,2}(t)>\Delta S_{A,1}(t)$ at long times beyond the saddle point approximation and matches condition (ii) of the main text.

\end{document}